\documentclass{PoS}
\newcommand{\beq}{\begin{equation}}
\newcommand{\eeq}{\end{equation}}
\newcommand{\bea}{\begin{eqnarray}}
\newcommand{\eea}{\end{eqnarray}}
\def\Id{ \mbox{1\hspace{-1.2mm}I} }

\def\BA{\begin{eqnarray}}
\def\EA{\end{eqnarray}}
\def\BAN{\begin{eqnarray*}}
\def\EAN{\end{eqnarray*}}

\def\tr{\mbox{tr}}

\def\u{{\bf u}}
\def\d{{\bf d}}
\def\s{{\bf s}}
\def\c{{\bf c}}
\def\b{{\bf b}}

\def\sbar{\bar{\bf s}}
\def\cbar{\bar{\bf c}}

\usepackage{graphicx}

\title{$B_s$ and $B_c$ mesons in lattice QCD with exact chiral symmetry}

\ShortTitle{$B_s$ and $B_c$ mesons in lattice QCD}

\author{TWQCD Collaboration: Ting-Wai Chiu$^a$, \speaker{Tung-Han Hsieh}$^b$ \\
\llap{$^a$}  Department of Physics and Center for Theoretical Sciences,
             National Taiwan University, Taipei~10617, Taiwan \\
             E-mail: \email{twchiu@phys.ntu.edu.tw} \\
\llap{$^b$}  Research Center for Applied Sciences, Academia Sinica,
             Taipei~115, Taiwan \\
             E-mail: \email{thhsieh@twcp1.phys.ntu.edu.tw}}  
\abstract{%
We determine the masses 
and decay constants of the pseudoscalar mesons $ B_s $ and $ B_c $, 
and also the masses of the vector mesons $ B_s^* $ and $ B_c^* $,  
in quenched lattice QCD with exact chiral symmetry.
For 100 gauge configurations generated with single-plaquette
action at $ \beta = 7.2 $ on the $ 32^3 \times 60 $ lattice,
we compute point-to-point quark propagators for 33 quark masses
in the range $ 0.01 \le m_q a \le 0.85 $, and measure the
time-correlation functions of pseudoscalar and vector mesons.
The inverse lattice spacing and the charm quark bare mass 
are determined using the mass and decay constant of $ \eta_c(2980) $. 
The bare masses of $ \s$ and $ \b $ quarks 
are chosen such that the masses of the corresponding vector mesons 
are in good agreement with $ \phi(1020) $,  
and $ \Upsilon(9460) $ respectively.
Our results are:
$ m_{B_s} = 5385(27)(17)$ MeV, 
$ f_{B_s} = 253(8)(7) $ MeV, 
$ m_{B_c} = 6278(6)(4)$ MeV, 
$ f_{B_c} = 489(4)(3)$ MeV, 
$ m_{B_s^*} = 5424(28)(19)$ MeV,
and $ m_{B_c^*} = 6315(6)(5)$ MeV.%
}

\FullConference{XXIVth International Symposium on Lattice Field Theory\\
                July 23-28, 2006\\
                Tucson, Arizona, USA}

\begin{document}

\section{Introduction}
 
In view of recent experimental results from B factories,
it is interesting to understand the B physics 
from the first principles, in the framework of lattice QCD 
with exact chiral symmetry. In order to be consistent with the 
Standard Model, here we treat {\it all} quark flavors (heavy and light) as 
Dirac fermions, without using any heavy quark and/or non-relativistic
approximations. Thus our approach is theoretically appealing,     
fundamentally different from other lattice QCD calculations
with heavy quark and/or non-relativistic 
approximations, in which the systematic errors are  
difficult to control. In spite of the large separation 
of mass scales for heavy-light systems, we can accommodate strange, 
charm and bottom quarks on a $ 32^3 \times 60 $ lattice, 
with inverse lattice spacing $ a^{-1} = 7.68 $ GeV. 
Then we compute point-to-point quark propagators,  
and extract physical quantities from the time-correlation function
of hadron interpolators containing $ \s $, $ \c $, and $ \b $ 
quarks. In this paper, we determine the masses 
and decay constants of the pseudoscalar mesons $ B_s $ and $ B_c $, 
and also the masses of the vector mesons $ B_s^* $ and $ B_c^* $. 
Note that $ B_c^* $ has not been observed in high energy experiments. 
Thus our result serves as the first prediction of $ B_c^* $
from lattice QCD.

To implement exact chiral symmetry on the lattice 
\cite{Kaplan:1992bt,Narayanan:1995gw,Neuberger:1997fp,Ginsparg:1981bj},
we consider the optimal domain-wall fermion proposed by 
Chiu \cite{Chiu:2002ir}. 
From the generating functional for $n$-point Green's function of the quark
fields, the valence quark propagator in background gauge field can be derived 
as \cite{Chiu:2002ir}
\bea
\label{eq:quark_prop}
\langle q(x) \bar q(y) \rangle 
&=& (D_c + m_q)^{-1}_{x,y}=(1-rm_q)^{-1}[D^{-1}_{x,y}(m_q) - r \delta_{x,y}]
\ , \hspace{5mm} r = (2 m_0)^{-1} \\
D(m_q) &=& m_q+(m_0-m_q/2) \left[1+\gamma_5 S(H_w) \right]  
\eea
where $m_q$ is the bare quark mass, $m_0$ is a parameter 
in the range $ (0,2) $, $S(H_w)$ is the Zolotarev approximation of the 
sign function of $H_w$ 
($H_w = \gamma_5 D_w$, and $D_w$ is the standard Wilson Dirac 
operator minus $m_0$), 
and $D_c = 2 m_0 (\Id+\gamma_5 S)(\Id-\gamma_5 S)^{-1} $, 
which becomes exactly chirally symmetric 
(i.e. $ D_c \gamma_5 + \gamma_5 D_c = 0 $) in the limit $ N_s \to \infty $ 
(where $ N_s + 2 $ is the number of sites in the 5th dimension.) 
Note that in this framework, the bare quark mass $ m_q $ 
in the valence quark propagator $ (D_c + m_q)^{-1} $ 
is well-defined for any gauge field configuration.   

In practice, there are two ways to evaluate the valence quark propagator 
(\ref{eq:quark_prop}): (i) To solve the linear system of the 5D 
optimal domain-wall fermion operator; 
(ii) To solve $ D(m_q) Z = \Id $ by nested conjugate gradient.  
Here we employ the scheme (ii), since we 
can attain the maximum efficiency 
if the inner conjugate gradient loop is iterated with
Neuberger's 2-pass algorithm \cite{Neuberger:1998jk}.

\begin{figure}[htb]
\begin{center}
\begin{tabular}{@{}cc@{}}
\includegraphics*[height=6.5cm,width=5.5cm]{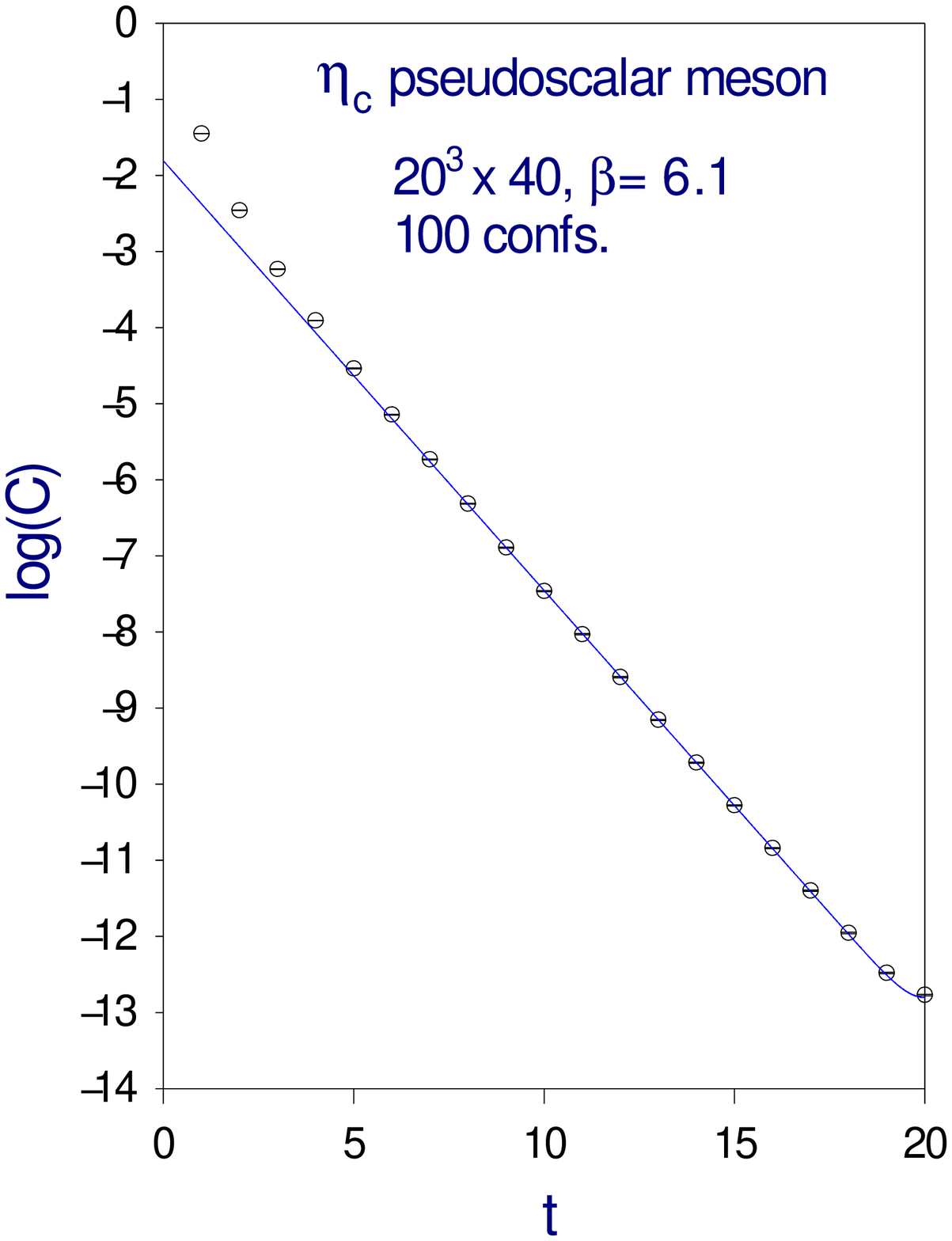}
&
\includegraphics*[height=6.5cm,width=5.5cm]{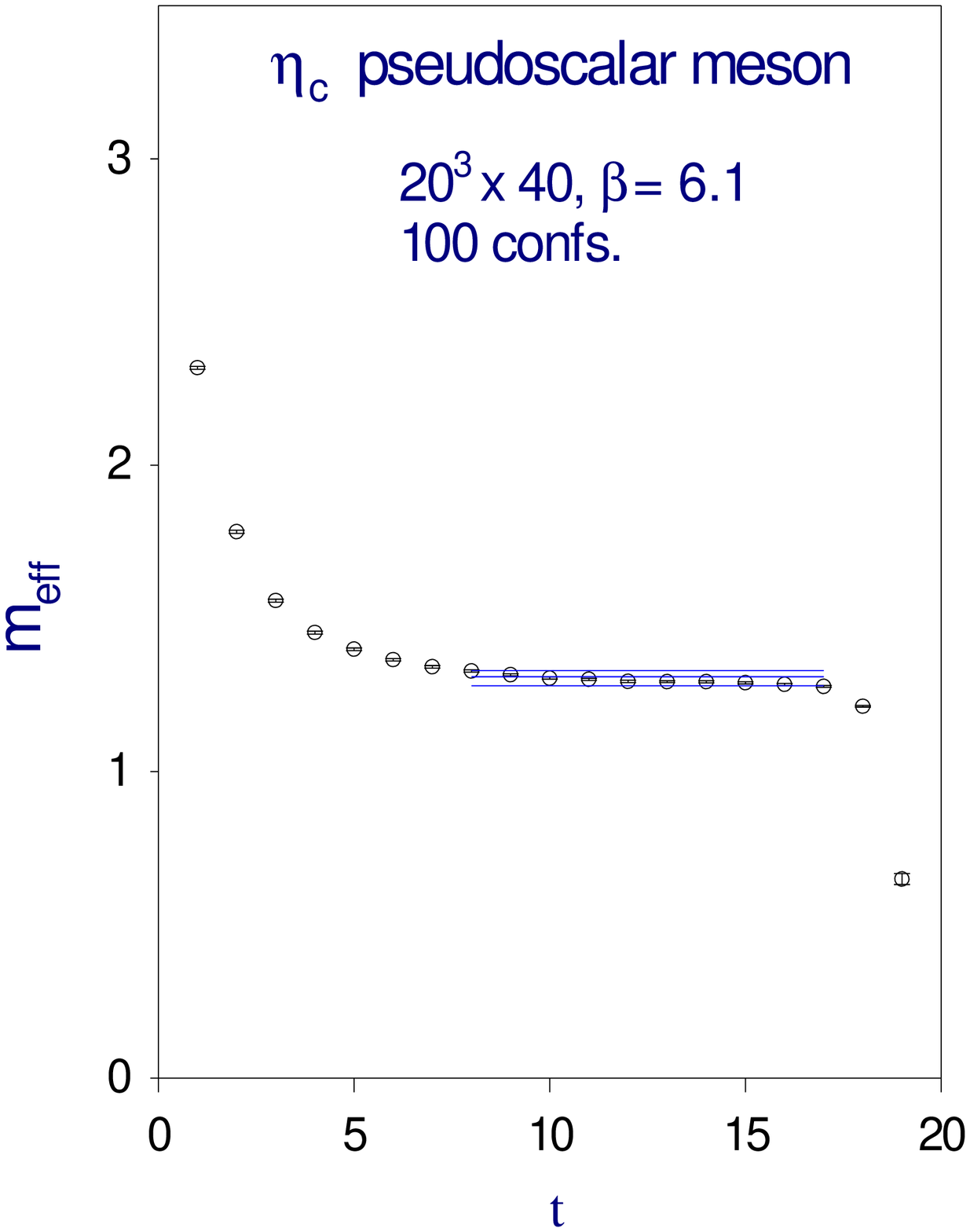}
\\ (a) & (b)
\end{tabular}
\caption{
(a) The time-correlation function $ C(t) $ of the
$ \eta_c $ pseudoscalar meson operator $ \cbar \gamma_5 \c $ 
on the $ 20^3 \times 40 $ lattice at $ \beta = 6.1 $.
The solid line is the hyperbolic-cosine fit for $ t \in [8,17] $.
(b) The effective mass $ M_{eff}(t) = \ln [C(t)/C(t+1)]  $
of $ C(t) $ in Fig.\ 1a.
}
\label{fig:etac}
\end{center}
\end{figure}

We generate 100 gauge configurations with single plaquette gauge action
at $ \beta = 7.2 $ on the $ 32^3 \times 60 $ lattice.
For $ m_0 = 1.3 $ and $ N_s = 128 $, 
we fix the Zolotarev coefficient with $ \lambda_{min} = 0.1 $
and $ \lambda_{max} = 6.4 $, 
where $ \lambda_{min} \le \lambda(|H_w|) \le \lambda_{max} $
for all gauge configurations.    
For each configuration, point-to-point quark propagators are computed
for 33 bare quark masses in the range $ 0.01 \le m_q a \le 0.85 $, 
with stopping criteria $ 10^{-11} $ and $ 2 \times 10^{-12} $
for the outer and inner conjugate gradient loops respectively.
Then the norm of the residual vector is 
$|| (D_c + m_q ) Y - \Id || < 2 \times 10^{-11} $, 
and the chiral symmetry breaking due to
finite $N_s$ is $ | \langle S^2 \rangle - 1 | < 10^{-14} $ 
for every iteration of the nested conjugate gradient.

In this paper, we measure the time-correlation functions 
for pseudoscalar ($P$) and vector ($V$) mesons,
\bea
\label{eq:CP}
C_{P} (t)  &=& 
\langle 
\sum_{\vec{x}}
\tr\{ \gamma_5 (D_c + m_Q)^{-1}_{x,0} \gamma_5 (D_c + m_q)^{-1}_{0,x} \} 
\rangle_U, \\
\label{eq:CV}
C_V (t) &=& \langle 
\frac{1}{3} \sum_{\mu=1}^3 \sum_{\vec{x}}
\tr\{ \gamma_\mu (D_c + m_Q)^{-1}_{x,0} \gamma_\mu
     (D_c + m_q)^{-1}_{0,x} \} \rangle_U
\eea
where the subscript $ U $ denotes averaging over gauge configurations. 
Here $ C_{P}(t) $ and $ C_V(t) $ are measured for the following
three categories: 
(i) Symmetric masses ($ m_Q = m_q $) for 33 quark masses;  
(ii) $ (m_Q, m_q) = (m_b, m_s) $; and 
(iii) $ (m_Q, m_q) = (m_b, m_c) $.


\begin{figure}[htb]
\begin{center}
\begin{tabular}{@{}cc@{}}
\includegraphics*[height=6.5cm,width=5.5cm]{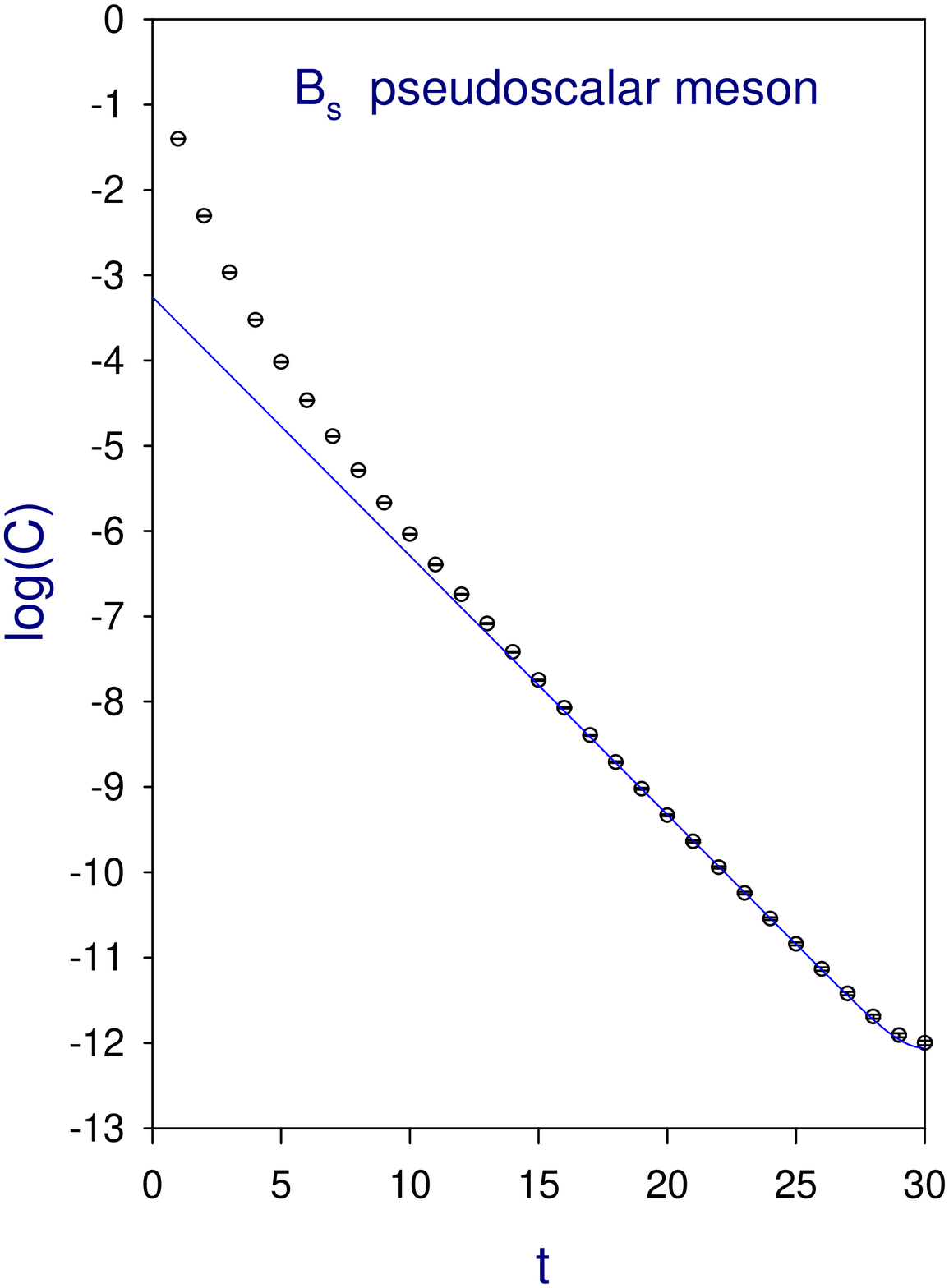}
&
\includegraphics*[height=6.5cm,width=5.5cm]{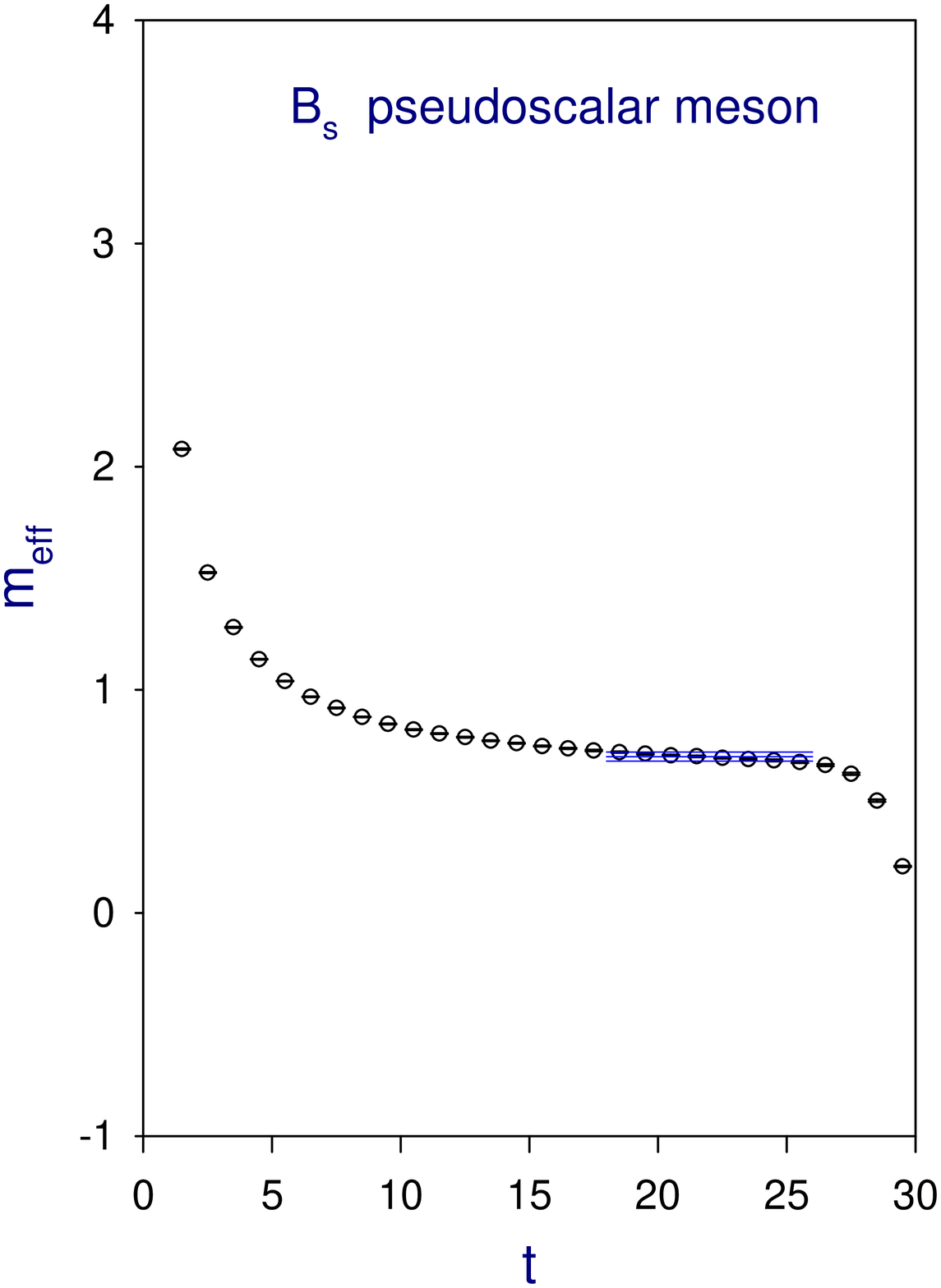}
\\ (a) & (b)
\end{tabular}
\caption{
(a) The time-correlation function $ C(t) $ of the
$ B_s $ pseudoscalar meson operator $ \sbar \gamma_5 \b $ with
$ J^{P} = 0^{-} $, on the $ 32^3 \times 60 $ lattice at $ \beta = 7.2 $.
The solid line is the hyperbolic-cosine fit for $ t \in [18,27] $.
(b) The effective mass $ M_{eff}(t) = \ln [C(t)/C(t+1)]  $
of $ C(t) $ in Fig.\ 2a.
}
\label{fig:Bs}
\end{center}
\end{figure}

\section{Determination of $ a^{-1} $, $ m_c $, $ m_s $, and $ m_b $}

In Ref.\ \cite{Chiu:2005ue}, we determine the inverse lattice 
spacing from the pion decay constant, with experimental input 
$ f_\pi = 131 $ MeV. However, in this paper, we do not use
the same method since the smallest quark mass   
turns out to be rather heavy ($ \simeq m_s/2 $), thus chiral 
extrapolation to $ m_q \simeq 0 $ does not seem to be feasible. 
Nevertheless, we can use the mass and decay constant of the 
pseudoscalar meson $ \eta_c(2980) $  
to determine $ m_c $ and $ a^{-1} $ simultaneously. 
This can be seen as follows.

For symmetric masses $ m_Q = m_q $, the pseudoscalar
time-correlation function $ C_{P} (t) $ (\ref{eq:CP}) is measured,
and fitted to the usual formula
\bea
\label{eq:Gt_fit}
\frac{z^2}{2 m_{P} a } [ e^{-m_{P} a t} + e^{-m_{P} a (T-t)} ]
\eea
to extract the mass $ m_{P} a $ and the decay constant
\bea
\label{eq:fpi}
f_{P} a = 2 m_q a \frac{z}{m_{P}^2 a^2 } \ .
\eea
Then the ratio $ m_{P}/f_{P} $ can be obtained for each $ m_q $.

From our previous studies \cite{Chiu:2005ue} of pseudoscalar mesons 
on the $ 20^3 \times 40 $ lattice at $ \beta = 6.1 $  
(see Fig. \ref{fig:etac}),   
we obtain the ratio $ m_{\eta_c}/f_{\eta_c} \simeq 6.8 $. 
Thus we can use this ratio to discriminate  
which $ m_q $ can give the ratio $ m_{P}/f_{P} $ closest to 6.8.
We find that at $ m_q a = 0.16 $, the ratio $ m_{P}/f_{P} = 6.8(1) $,
which is the closest to $ 6.8 $. Thus we fix $ m_c a = 0.16 $. 
Then we use the experimental mass of $ \eta_c(2980) $ to determine 
$ a^{-1} $ through the relation 
$$ 
m_{P} a|_{m_c} = (2980 \mbox{ MeV}) \times a = 0.388(3) 
$$
and obtain $ a^{-1} = 7680(59) $ MeV. 
To check the goodness of the values of $ m_c $ and $ a^{-1} $, 
we compute the time-correlation function of $ \cbar \gamma_i \c $, 
and extract the mass of the vector meson to be $ 3091(11) $ MeV, 
in good agreement with $ J/\Psi(3097) $. Note that the spatial 
size of our lattice ($ L \simeq 0.8 $ fm) seems to be small 
at first glance, however, even for the smallest quark mass 
$ m_q a = 0.01 $, its pseudoscalar mass satisfies $ m_{P} L > 4 $, 
thus the finite size effects are well under control.

\begin{figure}[htb]
\begin{center}
\begin{tabular}{@{}cc@{}}
\includegraphics*[height=6.5cm,width=5.5cm]{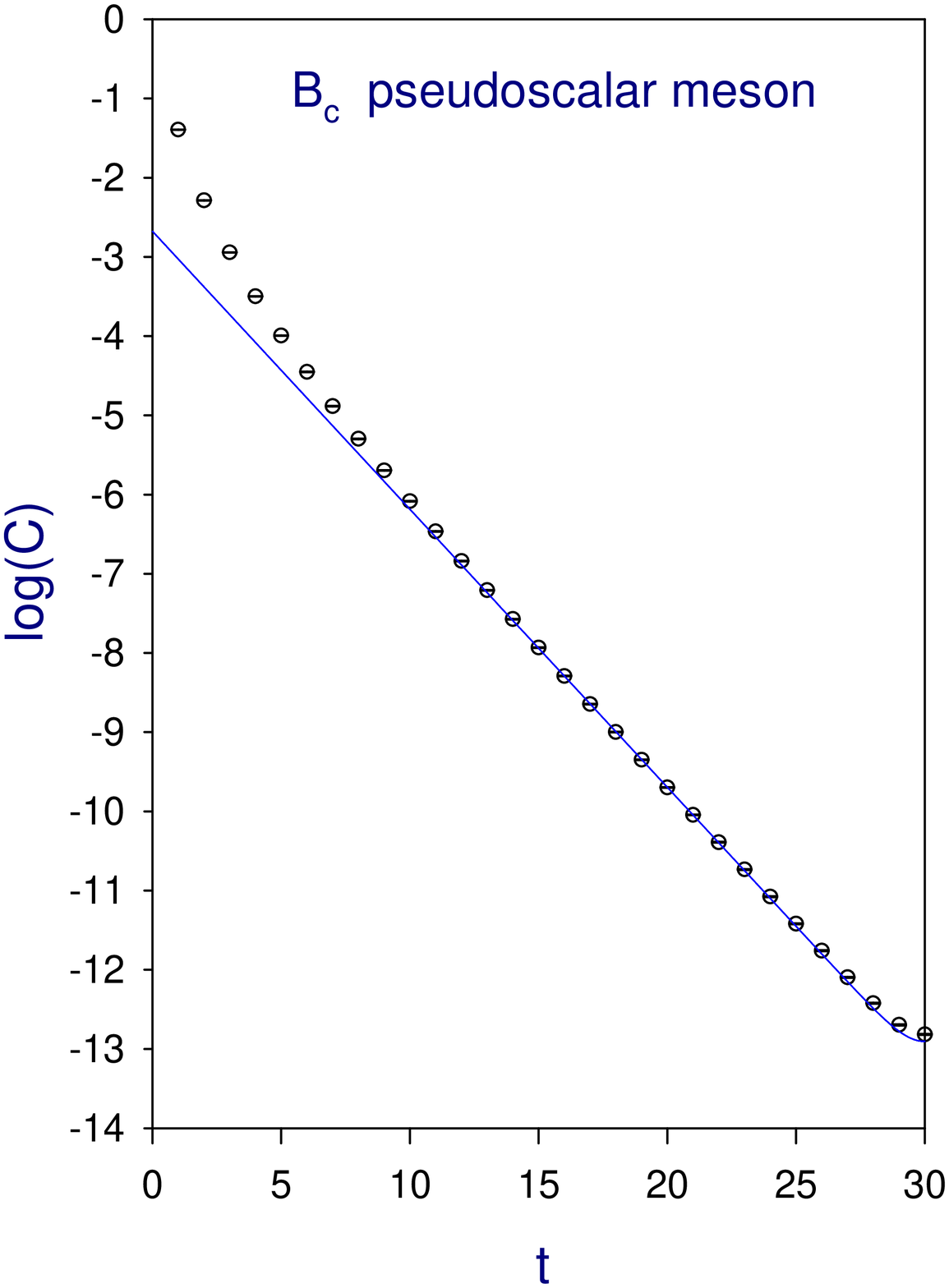}
&
\includegraphics*[height=6.5cm,width=5.5cm]{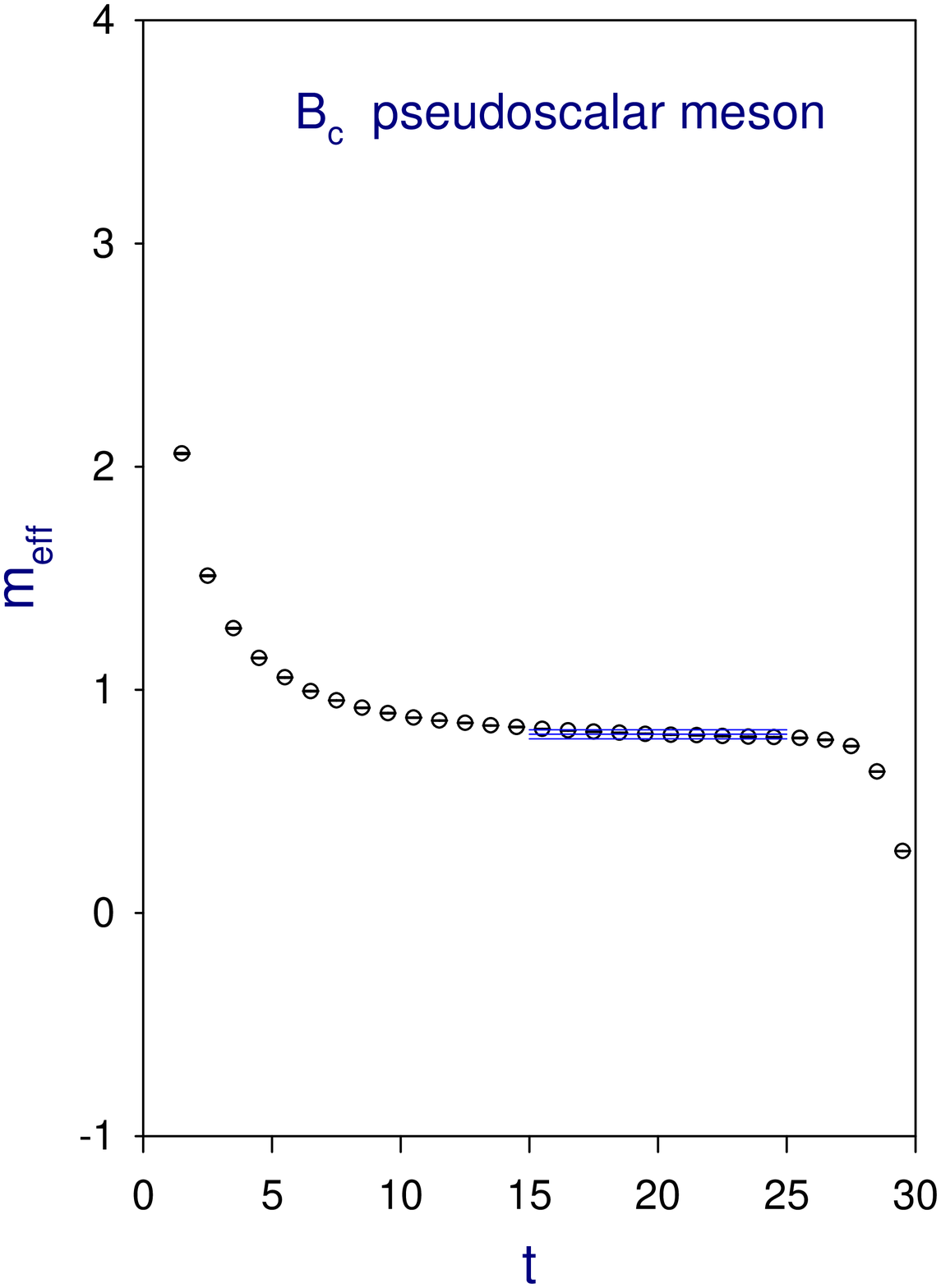}
\\ (a) & (b)
\end{tabular}
\caption{
(a) The time-correlation function $ C(t) $ of the
$ B_c $ pseudoscalar meson operator $ \cbar \gamma_5 \b $ with
$ J^{P} = 0^{-} $, on the $ 32^3 \times 60 $ lattice at $ \beta = 7.2 $.
The solid line is the hyperbolic-cosine fit for $ t \in [15,25] $.
(b) The effective mass $ M_{eff}(t) = \ln [C(t)/C(t+1)]  $
of $ C(t) $ in Fig.\ 3a.
}
\label{fig:Bc}
\end{center}
\end{figure}

The bare mass of strange quark is determined by extracting the
mass of vector meson from the time-correlation function
$ C_V(t) $.  At $ m_q a = 0.02 $, $ m_V a = 0.1337(5) $,
which gives $ m_V = 1027(38) $ MeV, in good agreement with
the mass of $ \phi(1020) $. Thus we take the strange quark
bare mass to be $ m_s a = 0.02 $.
Similarly, at $ m_q a = 0.68 $, $ m_V a = 1.2308(4) $,
which gives $ m_V = 9453(3) $ MeV, in
good agreement with the mass of $ \Upsilon(9460) $.
Thus, we fix the bottom quark bare mass to be $ m_b a = 0.68 $.

\section{The $ B_s $ and $ B_c $ Pseudoscalar Mesons}

The pseudoscalar-meson decay constants
play an important role in extracting the CKM matrix elements
which are crucial for testing the flavor sector of the
standard model via the unitarity of CKM matrix.
Theoretically, lattice QCD with exact chiral symmetry provides a 
reliable framework to compute the masses and decay constants of 
pseudoscalar mesons nonperturbatively
from the first principles of QCD.

The decay constant $ f_{P} $ for a pseudoscalar meson $ P $ is
defined by
\BAN
\left<0| A_\mu(0) | P(\vec{q}) \right> = f_P q_\mu
\EAN
where $ A_\mu = \bar q \gamma_\mu \gamma_5 Q $ is the axial-vector current. 
Using
$ \partial_\mu A_\mu = (m_q + m_Q) \bar q \gamma_5 Q $,
one obtains
\bea
\label{eq:fP}
f_P = (m_q + m_Q )
\frac{| \langle 0| \bar q \gamma_5 Q | P(\vec{0}) \rangle |}{m_P^2}
\eea
where the pseudoscalar mass $ m_P a $ and the decay amplitude
$ z \equiv | \langle 0| \bar q \gamma_5 Q | P(\vec{0}) \rangle | $
can be obtained by fitting the pseudoscalar time-correlation function
$ C_P(t) $ to the usual formula (\ref{eq:Gt_fit}).


\begin{figure}[htb]
\begin{center}
\begin{tabular}{@{}cc@{}}
\includegraphics*[height=6.5cm,width=5.5cm]{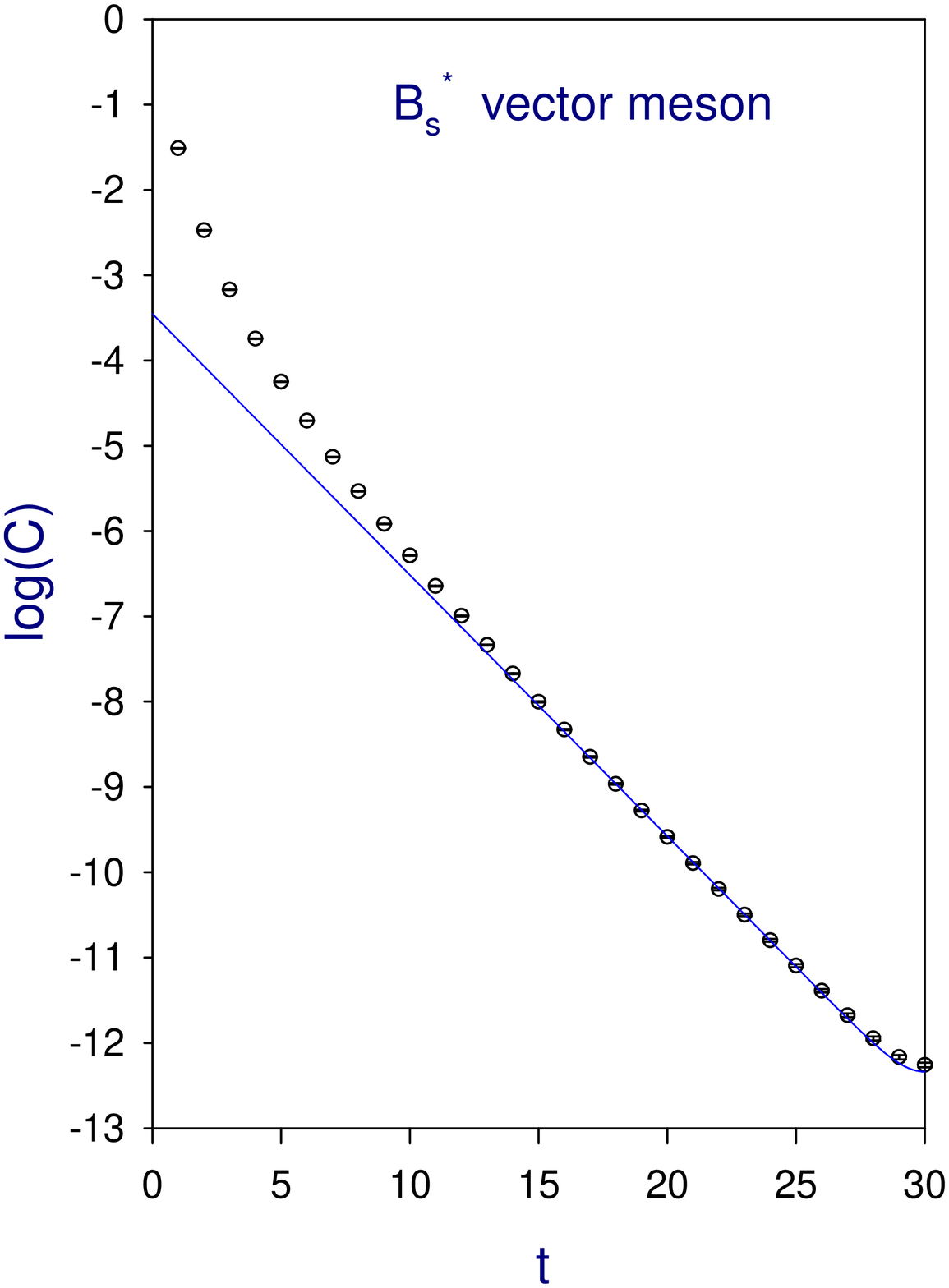}
&
\includegraphics*[height=6.5cm,width=5.5cm]{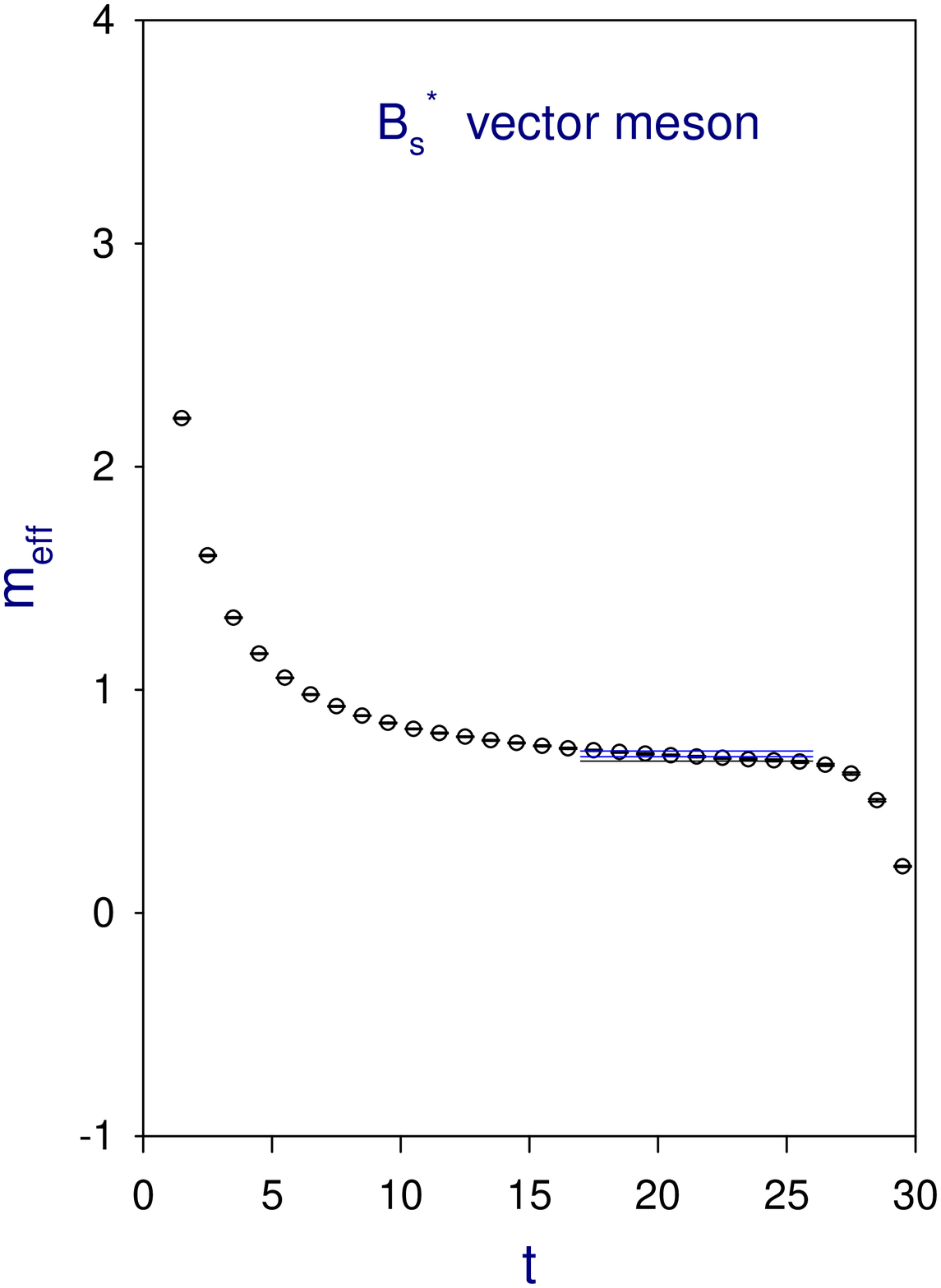}
\\ (a) & (b)
\end{tabular}
\caption{
(a) The time-correlation function $ C(t) $ of the
$ B_s^* $ vector meson operator $ \sbar \gamma_i \b $ with
$ J^{P} = 1^{-} $, on the $ 32^3 \times 60 $ lattice at $ \beta = 7.2 $.
The solid line is the hyperbolic-cosine fit for $ t \in [17,26] $.
(b) The effective mass $ M_{eff}(t) = \ln [C(t)/C(t+1)]  $
of $ C(t) $ in Fig.\ 4a.
}
\label{fig:BsV}
\end{center}
\end{figure}

In Fig.\ \ref{fig:Bs}, the time-correlation function $ C(t) $ 
of the $ B_s $ pseudoscalar meson operator $ \sbar \gamma_5 \b $ 
is plotted versus time slices, together with the effective mass
$ \log[C(t)/C(t+1)] $, for 100 gauge configurations generated 
with single plaquette action on $ 32^3 \times 60 $ lattice at 
$ \beta = 7.2 $. Here the ``forward-propagator" $ C(t) $ 
and the ``backward-propagator" $ C(T-t) $ are averaged to 
enhance the statistics. The same strategy is applied to all 
time-correlation functions in this paper.   
The solid line is the hyperbolic-cosine fit for $ t \in [18,27] $.
It gives $ m_{B_s} = 5385(27)(17)$ MeV and $ f_{B_s} = 253(8)(7) $ MeV,
where the first error is statistical, and the second is systematic
from all plausible fittings with $\chi^2/\mbox{dof} \le 1 $.
Evidently, $ m_{B_s} $ is in good agreement with the experimental 
value 5370 MeV. Since $ f_{B_s} $ has not been measured in high energy
experiments, our result serves as the first prediction from 
lattice QCD with exact chiral symmetry.

In Fig.\ \ref{fig:Bc}, the time-correlation function $ C(t) $ 
of the $ B_c $ pseudoscalar meson operator $ \cbar \gamma_5 \b $ 
is plotted versus time slices, together with the effective mass
$ \log[C(t)/C(t+1)] $. 
The solid line is the hyperbolic-cosine fit for $ t \in [15,25] $.
It gives $ m_{B_c} = 6278(6)(4)$ MeV, and $ f_{B_c} = 489(4)(3)$ MeV, 
where $ m_{B_c} $ is in good agreement with the experimental 
value 6287(5) MeV measured by CDF Collaboration. 
Since $ f_{B_c} $ has not been measured in high energy
experiments, our result serves as the first prediction from 
lattice QCD.

\section{The $ B_s^* $ and $ B_c^* $ Vector Mesons}

In Fig.\ \ref{fig:BsV}, the time-correlation function $ C(t) $ (\ref{eq:CV}) 
of the $ B_s^* $ vector meson operator $ \sbar \gamma_\mu \b $ 
is plotted versus time slices, together with the effective mass
$ \log[C(t)/C(t+1)] $. 
The solid line is the hyperbolic-cosine fit for $ t \in [17,26] $.
It gives $ m_{B_s^*} = 5424(28)(19)$ MeV, 
in good agreement with the experimental value 5417 MeV. 

\begin{figure}[htb]
\begin{center}
\begin{tabular}{@{}cc@{}}
\includegraphics*[height=6.5cm,width=5.5cm]{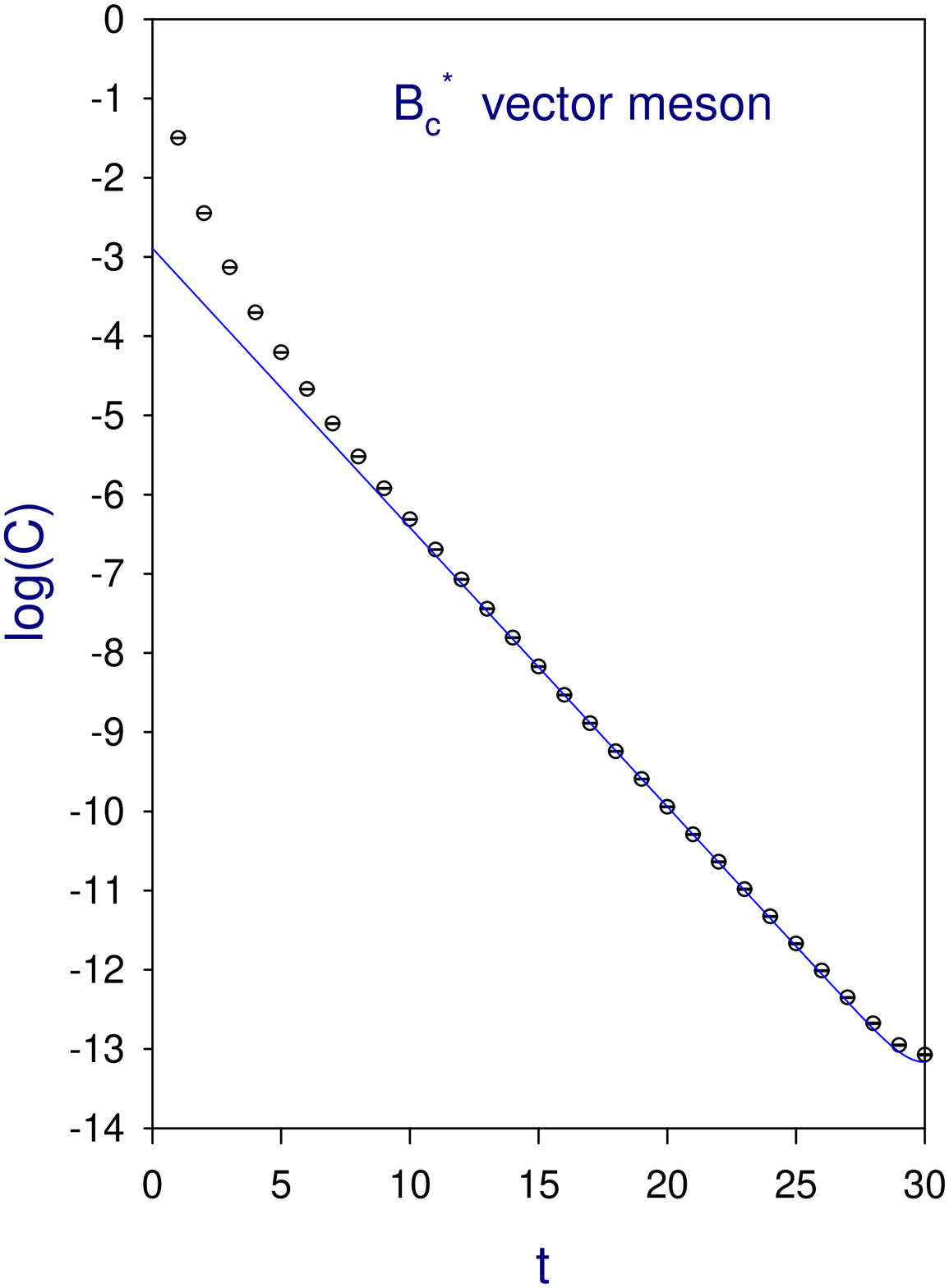}
&
\includegraphics*[height=6.5cm,width=5.5cm]{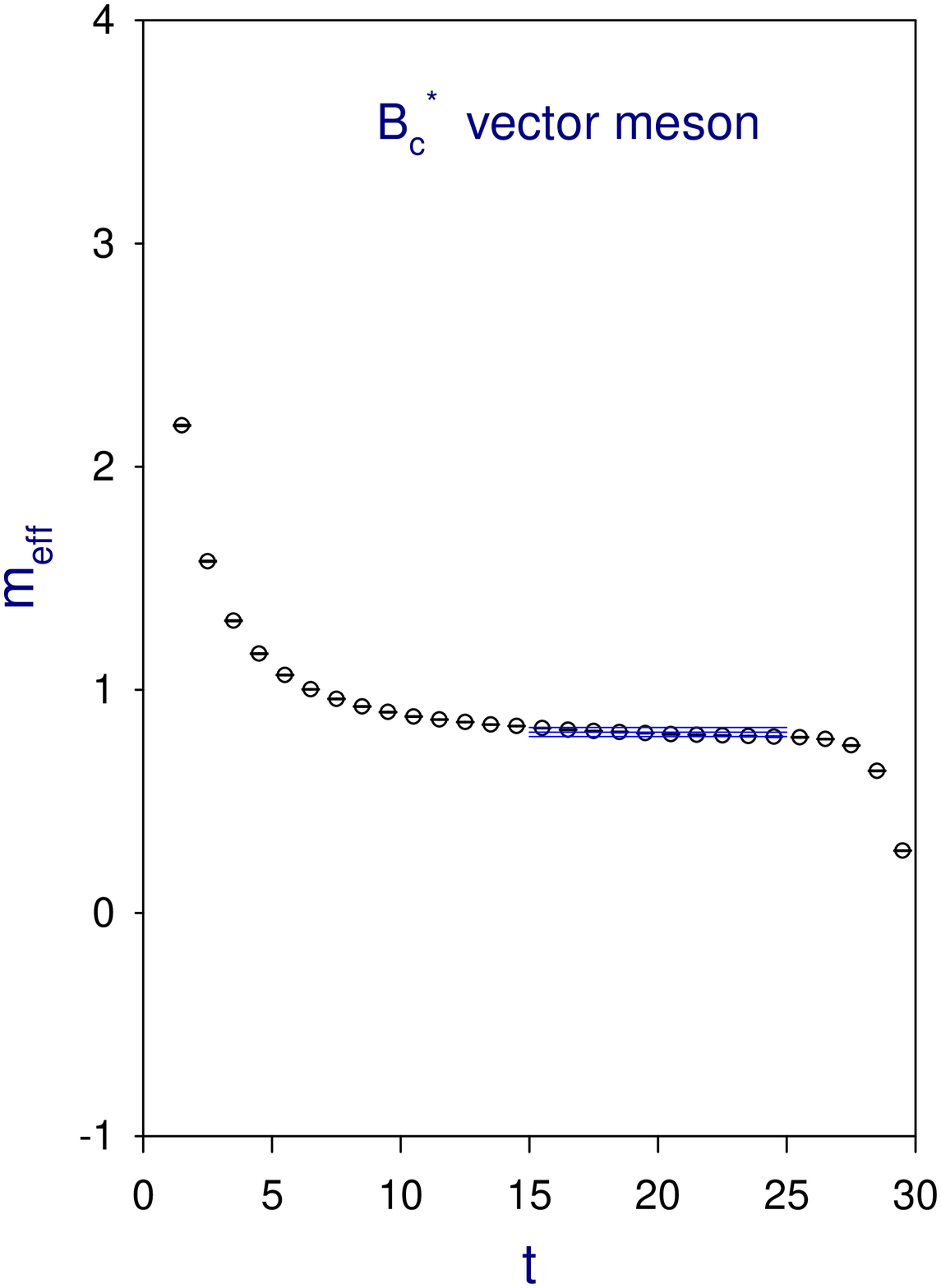}
\\ (a) & (b)
\end{tabular}
\caption{
(a) The time-correlation function $ C(t) $ of the
$ B_c^* $ vector meson operator $ \cbar \gamma_i \b $ with
$ J^{P} = 1^{-} $, on the $ 32^3 \times 60 $ lattice at $ \beta = 7.2 $.
The solid line is the hyperbolic-cosine fit for $ t \in [15,25] $.
(b) The effective mass $ M_{eff}(t) = \ln [C(t)/C(t+1)]  $
of $ C(t) $ in Fig.\ 5a.
}
\label{fig:BcV}
\end{center}
\end{figure}

In Fig.\ \ref{fig:BcV}, the time-correlation function $ C(t) $ (\ref{eq:CV}) 
of the $ B_c^* $ vector meson operator $ \cbar \gamma_\mu \b $ 
is plotted versus time slices, together with the effective mass
$ \log[C(t)/C(t+1)] $. 
The solid line is the hyperbolic-cosine fit for $ t \in [15,25] $.
It gives $ m_{B_c^*} = 6315(6)(5)$ MeV, which serves as 
the first prediction from lattice QCD.

\section{Concluding Remark}

In this paper, we have investigated  
heavy quark systems containing $ \b $, $ \c $ and $ \s $, 
treating all quarks as Dirac fermions on the lattice, 
without using any heavy quark and/or non-relativistic approximations. 
Our results of the masses and decay constants of the pseudoscalar mesons 
$ B_s $ and $ B_c $, and also the masses of the vector mesons 
$ B_s^* $ and $ B_c^* $ suggest that lattice QCD with exact chiral 
symmetry is a viable framework to study heavy quark physics 
from the first principles of QCD.  
For systems involving $ \u/\d $ quarks, one may use several quark masses 
in the range $ m_{u/d} < m_q < m_{s} $ to perform the chiral extrapolation. 
To this end, one may choose a coarser lattice (e.g. $ \beta = 7.0 $),  
then it is possible to accommodate a wide range of quark masses  
$ m_s/4 < m_q \le m_b $ on the $ 42^3 \times 64 $ lattice, 
without significant discretization and finite-size errors.  
Obviously, it has become feasible to 
treat all quarks as Dirac fermions on the lattice, 
in lattice QCD with exact chiral symmetry.

\section*{Acknowledgement}

This work was supported in part by the National Science Council,
Republic of China, under the Grant No. NSC95-2112-M002-005 (T.W.C.),  
and Grant No. NSC95-2112-M001-072 (T.H.H.), and by the
National Center for High Performance Computation at Hsinchu, 
and the Computer Center at National Taiwan University.

\end{document}